\documentclass{article}

\usepackage{amsmath}
\usepackage{PRIMEarxiv}
\usepackage[numbers]{natbib}
\usepackage[utf8]{inputenc} 
\usepackage[T1]{fontenc}    
\usepackage{hyperref}       
\usepackage{url}            
\usepackage{booktabs}       
\usepackage{amsfonts}       
\usepackage{nicefrac}       
\usepackage{microtype}      
\usepackage{lipsum}
\usepackage{fancyhdr}       
\usepackage{graphicx}       
\usepackage{amsmath}
\usepackage{xcolor}
\usepackage{svg}
\usepackage{appendix}
\usepackage{longtable}
\usepackage{tabularx}
\usepackage{amsmath}
\usepackage{subcaption}
\graphicspath{{media/}}     

\title{Learning Electromagnetic Metamaterial Physics
with ChatGPT
\thanks{\textit{\underline{Citation}}: 
\textbf{Authors. Title. Pages.... DOI:000000/11111.}} 
}

\author{
  Darui Lu, Yang Deng, Willie J. Padilla* \\
  Pratt School of Engineering\\
  Duke University \\
  Durham, North Carolina, 27705, USA\\
  \texttt{*willie.padilla@duke.edu} \\
    \And
  Jordan M. Malof \\
  Department of Computer Science \\
  University of Montana \\
  Missoula, Montana 59812, USA \\
}

\begin{document}
\maketitle

\begin{abstract}
Large language models (LLMs) such as ChatGPT, Gemini, LlaMa, and Claude are trained on massive quantities of text parsed from the internet and have shown a remarkable ability to respond to complex prompts in a manner often indistinguishable from humans. For all-dielectric metamaterials consisting of unit cells with four elliptical resonators, we present a LLM fine-tuned on up to 40,000 data that can predict the absorptivity spectrum given a text prompt that only specifies the metasurface geometry. Results are compared to conventional machine learning approaches including feed-forward neural networks, random forest, linear regression, and K-nearest neighbor (KNN). Remarkably, the fine-tuned LLM (FT-LLM) achieves a comparable performance across large dataset sizes with a deep neural network. We also explore inverse problems by asking the LLM to predict the geometry necessary to achieve a desired spectrum. LLMs possess several advantages over humans that may give them benefits for research, including the ability to process enormous amounts of data, find hidden patterns in data, and operate in higher-dimensional spaces. This suggests they may be able to leverage their general knowledge of the world to learn faster from training data than traditional models, making them valuable tools for research and analysis.
\end{abstract}



\section{Introduction}
\label{sec:introduction}
Deep learning, and particularly deep neural networks (DNNs), have recently emerged as a valuable tool in the field of metamaterials research and have produced many novel results\cite{ma2021deep, jiang2021deep, khatib2021deep}. This data-driven approach to metamaterial design has profound capabilities for both forward \cite{yao2019intelligent} and inverse processes\cite{so2020deep, wiecha2021deep}. Once trained, DNNs can accelerate the simulation of material systems by orders of magnitude compared to traditional numerical simulations \cite{peurifoy2018nanophotonic, ma2018deep, jiang2019global, nadell2019deep}, enabling faster prototyping and exploration. Similarly, in inverse design, these models have successfully discovered state-of-the-art solutions that push the boundaries of what is achievable with metamaterials \cite{liu2018generative, jiang2019free, deng2021neural, ren2022inverse}.

Despite these advances, the implementation of DNNs still faces several challenges \cite{ma2021deep, jiang2021deep, khatib2021deep}. As a data-driven method, DNNs necessitate large datasets for training to achieve high accuracy and generalizability \cite{lecun2015deep}. This so-called "data bottleneck" issue is compounded by questions of interpretability and understanding, and therefore explaining model predictions remains a significant hurdle\cite{zhang2021understanding, rudin2022interpretable}.This has led to the pursuit of models that are capable of learning effectively from smaller datasets, and to the development of techniques such as transfer learning, \cite{xu2021efficient, zhu2021phase, zhang2022heterogeneous, fan2022transfer} and physics-informed / driven DNNs  \cite{raissi2019physics, jiang2019global, chen2020physics, chen2022high, khatib2022learning, ji2023recent, augenstein2023neural}.  

Foundational models are sophisticated, large-scale DNNs trained on extensive, diverse datasets \cite{fomos2021}. This training enables them to generalize knowledge across various domains without the need for substantial task-specific data \cite{fomos2021, brown2020language}. However, the significant cost of acquiring large, diverse datasets for training a physics foundational model has thus far been prohibitive for academic researchers \cite{mccabe2023multiple}. In this work, we hypothesize that LLMs, such as ChatGPT, can leverage their broad capabilities to develop a deeper understanding of physical systems, going beyond predictive capabilities, with far less training data than existing deep neural network (DNN) models. Rather than building a foundational physics model from scratch, we explore re-purposing existing foundational models -- also termed large language models (LLMs) -- to address problems in metamaterial design, and show promising results \cite{brown2020language,touvron2023llama}.

LLMs, such as the widely-used generative pre-trained transformer (GPT) models, have recently emerged as a foundational model primarily designed to handle natural language processing tasks \cite{brown2020language}. These models are trained to perform a simple task: given some textual input (e.g., a sentence fragment), predict the most probable next word of the input.  Even though this training task is relatively simple, the models are trained on massive volumes of textual data, and they have been found to acquire an ability to construct coherent and contextually relevant text that encapsulates broad knowledge across diverse domains.  There is also some evidence that LLMs can utilize this apparent knowledge to perform basic reasoning tasks \cite{vaswani2017attention}. For instance, LLMs can engage in conversations, translate languages, summarize texts, and even generate content that mimics human writing styles \cite{brown2020language,touvron2023llama}.

The multifaceted capabilities of LLMs are rooted in their extensive training on diverse datasets \cite{chang2023survey}. First, they integrate a vast range of information from their training sets, making them repositories of wide-reaching knowledge about much the world. This allows them to recall and leverage facts, concepts, and relationships when generating responses. Secondly, LLMs can perform reasoning tasks based on the information they have been trained on, enabling them to handle queries that require logical deductions, problem-solving, or creative generation. This can range from solving mathematical problems to crafting detailed narratives or technical explanations. Lastly, the ability of LLMs to explain their reasoning process adds a layer of interpretability, often allowing users to understand the steps the model took to arrive at a conclusion or response, thus providing insights into the model’s thought process.

In the case of metamaterial design, LLMs such as ChatGPT could revolutionize how physical systems are modeled and understood with minimal training data \cite{hu2024language}. The diverse and extensive textual training of LLMs encompasses fundamental physics concepts, which are crucial for understanding the dynamics of metamaterials. Since LLMs have substantial internal physics knowledge, they could leverage this knowledge to perform basic reasoning about such systems. For example, they may be able to accurately predict the behavior of physical systems using far less empirical training data than might be required when training DNNs from scratch, as is often done now.  Furthermore, LLMs may also be able to explain the reasoning that leads to their predictions using  natural language.  If feasible, LLMs utilized in this manner could significantly accelerate the design and simulation processes in metamaterial engineering.

Our study investigates the feasibility of LLMs to address metamaterial design challenges. Building on recent findings that suggest LLM proficiency in scientific regression and classification tasks \cite{dinh2022lift}, we explore their potential in predicting the electromagnetic spectra of metamaterials -- a problem expressible in textual terms. This adaptability of LLMs has already been demonstrated in chemistry \cite{jablonka2024leveraging, jablonka202314}, optics \cite{ma2023optoGPT}, and mechanics\cite{buehler2023melm}, signaling their versatility across various scientific fields. 

Our investigation focuses on all-dielectric metasurfaces, including prediction of the absorptivity spectrum, inverse design, and extracting insights. We compare the capabilities of LLMs with established machine learning and deep learning models. To our knowledge, there has yet to be a study of the use of LLMs on regression tasks with high-dimensional outputs, such as those often encountered in metamaterial problems. 

We find that our fine-tuned model, FT-LLM (based on ChatGPT 3.5), achieves a competitive performance with deep neural networks on large datasets for prediction of the frequency dependent absorptance. Moreover, one of the most powerful capabilities of LLMs is their ability to summarize high-dimensional or voluminous data with interpretable natural language, such as generating short captions that succinctly describe image content \cite{NEURIPS2023_6dcf277e}. Motivated by this, we investigate whether LLMs might be able to summarize insights from large datasets of simulation data. In particular, we explore whether our FT-LLM model can explain what it has learned about all-dielectric metasurfaces (ADMs), and whether these explanations are superior to those obtained by an out-of-the-box LLM that has not been trained on our data. Additionally, we assess the capabilities of the FT-LLM to perform inverse design by evaluating whether it can generate metamaterial geometries that achieve specified absorptivity spectra. Through this work, we highlight the potential of LLMs as powerful tools in scientific exploration, potentially broadening the horizons for future innovations and discoveries in metamaterials and beyond.

\section{Methodology}
\label{sec: methodology}

This work adapts LLMs to predict the spectral response of metamaterials. Specifically, we focus on predicting the spectra of an all-dielectric metamaterial, consisting of a unit cell with four elliptical resonators characterized by fourteen geometric parameters: height ($h$), periodicity ($p$), semimajor axes ($r_{ma}$), semiminor axes ($r_{mi}$), and rotational angles ($\theta$). Let \(\mathcal{D} = \{ (g_{(i)}, s_{(i)}) \mid i = 1, \dots, n \}\) be our simulation dataset\cite{deng2021neural}. Here, simulation data are generated through use of a computational electromagnetic solver, CST Studio Suite (CST)\cite{studio2010cst}. Thus the simulated data constitute the "ground truth" data. Given a raw geometry vector \( g_{(i)}  \in \mathbb{R}^{14}\) , our objective is to predict its corresponding spectrum \( s_{(i)} \in \mathbb{R}^{50} \). However, LLMs, represented here as \( f_{\phi} \) with trainable parameter $\phi$ , are designed to process text data. With an text input, the model produces an output text completion.

\begin{figure*}[!ht]
     \centering
     \includegraphics[width=0.75\textwidth]{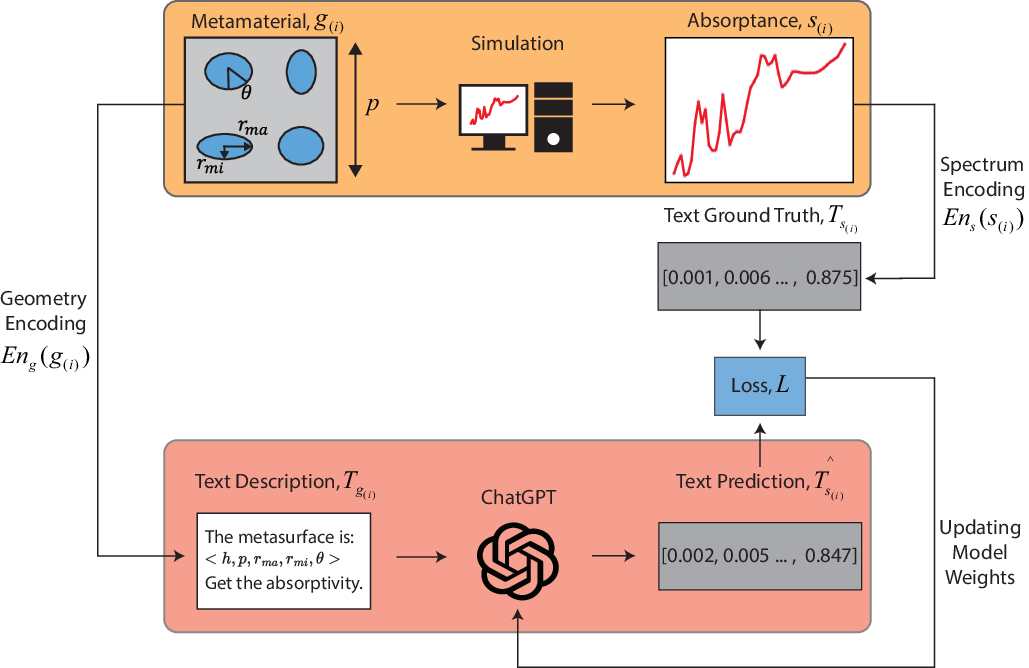}
     \caption{Schematic depicting the workflow. The process begins with simulations (orange rectangle) to acquire the metamaterial geometry-spectrum dataset. Subsequently, these numerical geometric parameters are transformed into textual descriptions -- shown by the numerical encoding path. We then fine-tune ChatGPT 3.5 using OpenAI's API with the built-in loss shown as the blue box.}
     \label{fig: convert}
\end{figure*}

LLMs currently do not possess the ability to directly process numerical data \( g_{(i)} \) and \( s_{(i)}\). To address this issue, we transform the tabular data into text descriptions \( T_{g{(i)}}=En_g(g_{(i)}) \) and \( T_{s{(i)}}=En_s(s_{(i)}) \), where \( En_g \) and \( En_s \) are transformation functions that converts numerical data into a structured text format. Based upon the textual input \( T_{g{(i)}} \), the LLMs can then produce a textual completion, \( \hat{T}_{s{(i)}} \) = \( f_{\phi} (T_{g{(i)}})\).  We can then fine-tune a LLM to make more accurate predictions by comparing its output to $T_{s{(i)}}$. As a result, we enable the model to handle the spectral response prediction as a language processing task without any modifications to the model structure and training process \cite{dinh2022lift}. The flowchart in Fig. \ref{fig: convert} illustrates the steps of our method.

\subsection{Numerical Encoding}

Our first step is to transform the numerical data into a suitable format for input to the LLM. For the numerical vector describing the geometry of the metasurface, we embed it within a text prompt template \cite{buehler2023melm}. Given a raw geometry vector \(g_{(i)} = \) ($h$, $p$, $\text{r}_{ma1}$, $\text{r}_{mi1}$, $\text{r}_{ma2}$, $\text{r}_{mi2}$, $\text{r}_{ma3}$,$\text{r}_{mi3}$,$\text{r}_{ma4}$, $\text{r}_{mi4}$, $\theta_1$, $\theta_2$, $\theta_3$, $\theta_4$), the  text encoding \(T_{g{(i)}}\) is as follows: "The All-dielectric metasurface suspended in free space is: <$h$, $p$, $\text{r}_{ma1}$, $\text{r}_{mi1}$, $\text{r}_{ma2}$, $\text{r}_{mi2}$, $\text{r}_{ma3}$, $\text{r}_{mi3}$, $\text{r}_{ma4}$, $\text{r}_{mi4}$, $\theta_1$, $\theta_2$, $\theta_3$, $\theta_4$> Get the absorptivity". An example is provided in Table \ref{tab:Table1}.
 
In terms of spectra, we directly convert the ground truth \(s_{(i)}\) into strings \(T_{s{(i)}}\), such as translating number $0.1$ to '0.1'. An issue that arises is that the spectrum value can possess many digits, such as $0.010290466248989103$. Using a LLM to deal with a number such as this would be inefficient and -- more importantly --  challenging. LLMs are generative models and it is difficult for them to maintain prediction precision beyond a few decimal points \cite{jablonka2024leveraging}. Furthermore, error measurements are often insensitive to such precision, indicating that maintaining this level of significant digits is unnecessary. Therefore, we use a precision of three decimal points, generating $T_{s{(i)}}$ such as '[$0.001$, $\cdots$, $0.875$]'.

 \subsection{Model Fine-Tuning}
 Once the geometry-spectrum data has been translated into text, it can be used to fine-tune the LLM. In this work, we train the Generative Pre-rained Transformer 3.5 (ChatGPT 3.5) utilizing the OpenAI API. During the training process, the model receives \(T_{g{(i)}}\) and then produces a text response \( \hat{T}_{{s{(i)}}} \). By comparing prediction \(\hat{T_{s(i)}}\) and ground truth \(T_{s{(i)}}\), we iteratively adjust the model parameters $\phi$ to improve its spectral prediction capabilities. We call the resulting trained GPT model the Fine-Tuned LLM (FT-LLM). 

 \subsection{Inference}
 The temperature setting in LLMs influences the level of randomness in the generated results \cite{huang2023survey}. The formula is given by:
\begin{equation}
\text{P}(x_i) = \frac{e^{\left(\frac{1}{T}\text{logit}(x_i)\right)}}{\sum_{j} e^{\left(\frac{1}{T}\text{logit}(x_j)\right)}}
\end{equation}

\noindent where $\text{P}(x_i)$ is the probability of the $i$-th token, and $\text{logit}(x_i)=\ln(x_i/(1-x_i))$ for $x_i \in (0,1)$ is the logit (log probability) of the $i$-th token from the output layer and $T$ is the temperature.

We set the temperature at a value of $0.5$. Let \(\mathcal{D}_t = \{ (g_{t{(i)}}, s_{t{(i)}}) \mid i = 1, \dots, n \}\) be our test dataset. During inference (i.e., making predictions),  FT-LLM predicts a spectral response \(\hat{T}_{st(i)}\) for geometry input $T_{gt{(i)}}$. We then convert \(\hat{T}_{st{(i)}}\) back into numerical form $\hat{s_{t(i)}}$, which is then benchmarked against spectra generated by CST $s_{t(i)}$ to measure the accuracy. Since the FT-LLM can give undesired outputs, we further discuss error management in Appendix \ref{A: Managing output errors}.

\section{Experimental Design and Resources}
\label{Experimental}


\textbf{Dataset}
Our study harnesses a dataset introduced and benchmarked in previous work, \cite{deng2021neural, deng2021benchmarking} and is chosen for its relevance to the understanding of the capability of LLMs in designing metasurfaces. This dataset is open-access, and the structured vector format of its geometrical inputs and spectral outputs are advantages for prompt engineering in LLMs. All simulation data are generated using the CST solver. That is we find that the vector description provides a compact representation of the complex metasurface, with each parameter encoding its physical impact on absorption. Such datasets are helpful for physically data manipulation and model training, enabling a higher degree of interpretability of feature importance. Below, we detail the metasurface geometry and spectra in the dataset. 

The all-dielectric metasurface is fashioned from silicon carbide and operates over a range in the THz from 150-500 THz. This complex structure is defined by a supercell comprising four elliptical resonators, each positioned at the center of a subdivided quadrant within a square supercell. The metasurface geometry is given as a 14-dimensional vector: [$h$, $p$, $r_{ma1}$, $r_{mi1}$, $r_{ma2}$, $r_{mi2}$, $r_{ma3}$,$ r_{mi3}$, $r_{ma4}$, $r_{mi4}$, $\theta_1$, $\theta_2$, $\theta_3$, $\theta_4$]. The periodicity $p$ parameter specifies the side length of the supercell, and is proportional to the operating range of the resonator array. The height $h$ parameter is set to be identical for all resonators, and all geometrical parameters are in units of $\mu m$. The in-plane dimensions of each elliptical resonator are proportionally scaled to the supercell periodicity through the x-axis radius ratio $r_{ma,i}$ and y-axis radius ratio $r_{mi,i}$ for each resonator, respectively. Additionally, the orientation of each elliptical resonator is adjusted through a rotational angle $\theta_i$, measured in radians, with respect to the x-axis. Given the challenges associated with processing high-dimensional data by LLMs, the spectrum output was manipulated by first downsampling from 2000 frequency points to 100 frequency points. Then, we only select 50 points from the $150-350$ THz frequency range, aiming to refine the LLM's predictive accuracy and computational efficiency within the expansive operational bandwidth.

\textbf{Data Handling} To ensure experimental integrity, we divided the dataset into three distinct, independently sampled sets: training, validation, and test. The training set was randomly selected for each dataset size's training session, whereas the validation and test sets comprising 1,000 samples each were specified prior to the experiments. The test set was constructed to evaluate the model’s generalization performance, and it contains metamaterial geometries and absorptance values that the model has not seen during training or validation. Optimal model selection was based on validation set performance, with the test set used for final evaluation. This approach guarantees the reliability and reproducibility of our comparison between the baseline models and large language models applied to the metamaterials problem.

Prior to model training, our dataset undergoes preprocessing steps to ensure optimal model performance. The geometry inputs are normalized to a range of $[-1,1]$, promoting convergence in the NN training process. The absorptivity spectra are already within the range of $[0,1]$, and therefore require no additional preprocessing. Details regarding hyperparameter tuning are shown in Appendix \ref{A:Hyperpara}.

\textbf{Scoring Metrics}
\label{sec:experimental_design_scoring_metrics}
This section outlines the error metrics employed for evaluating the performance of both baseline and LLM models. Our models predict the electromagnetic spectra at a set of specific frequency values, and we wish to measure the error between these predictions and the CST-generated ground truth. In artificial electromagnetic material (AEM) research, both absolute and relative error metrics are commonly utilized \cite{peurifoy2018nanophotonic, nadell2019deep}. To comprehensively evaluate accuracy of the FT-LLM, we employ two complementary metrics: mean squared error (MSE) and mean absolute relative error (MARE). The MSE and MARE over the testing dataset are given mathematically by

\begin{equation}
\text{MSE} = \frac{1}{n}\sum_{i=1}^{n}\frac{1}{f}\sum_{j=1}^{f}(s_{i,j} - \hat{s}_{i,j})^2),
\end{equation}

\begin{equation}
\text{MARE} = \frac{1}{n}\sum_{i=1}^{n}\frac{1}{f}\sum_{j=1}^{f}\frac{|s_{i,j} - \hat{s}_{i,j}|}{|s_{i,j}|+\epsilon}),
\end{equation}

\noindent where $n$ is the number of samples, $f$ is the number of frequency points in the spectrum, $S_{i,j}$ represents the absorptivity value for $i$th sample at the $j$th frequency point, and $\hat{S}_{i,j}$ denotes the model predicted value. For MARE, we add a small value $\epsilon=1e^{-4}$ to mitigate the effect of tiny values which can lead to extremely large percentage errors, thereby skewing the metric -- see Appendix \ref{MARE}.

\textbf{Error Metric Bias} It is important to note that MSE is favorably biased towards all of the baseline models, since they are all trained using this metric. Therefore, the MSE gives a unfavorable bias for the FT-LLM giving it a disadvantage. However, the MARE provides a more unbiased metic for all models, since none of them are trained on this metric. Therefore, the baseline models will tend to perform better on MSE than MARE, as MARE is more sensitive to errors in low-value regions.  Additionally, LLMs are most often trained using cross-entropy loss (CEL)\cite{brown2020language}, which treats all errors equally, regardless of their magnitude, and therefore is not typically used to evaluate the accuracy of regression models. Because of this, and because CEL has not been utilized to evaluate the accuracy of such models in AEM studies, we do not include it in our evaluations.






\textbf{Baseline Models} To benchmark the performance of LLMs, we incorporate four other machine learning algorithms: Feed-forward Neural Networks (NN), Random Forests (RF), K Nearest Neighbors (KNN), and Linear Regression (LR). The selection of the NN was motivated by its published efficacy in various practical applications, particularly within the domain of metamaterials research, where its ability to model complex, nonlinear relationships is highly valued \cite{ma2021deep, jiang2021deep, khatib2021deep}. Conversely, RF, KNN, and LR represent classical machine learning algorithms designed to address regression challenges. Given that the design of metamaterials is predominantly formed as a regression problem, these algorithms were deemed especially suitable for benchmarking against the metasurface design challenges alongside LLMs. Our choice of algorithms aims to encompass both the cutting-edge capabilities of neural networks and the robust, well-established methodologies of classical machine learning, ensuring a comprehensive and fair evaluation of FT-LLM applicability and performance in the all-dielectric metasurface design.

\begin{figure*}[ptb]
     \centering
     \includegraphics[width=0.75\textwidth]{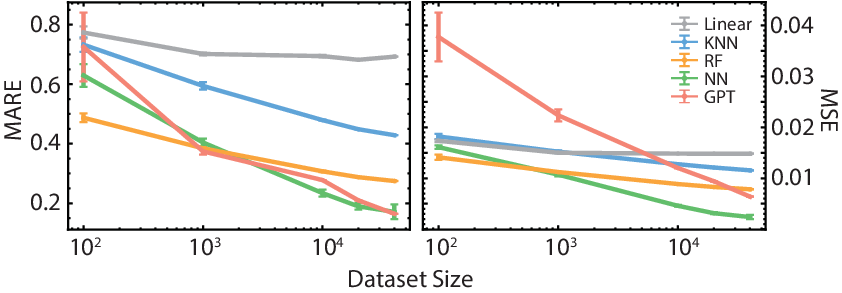}
     \caption{Evaluations of the models for predicting metamaterial spectra, given a geometry, as a function of dataset size. (a) MARE and (b) MSE trends for baseline models and the FT-LLM as dataset size increases. All results of baseline models presented are the average of three models. Error bars indicate the standard deviation of the three trails. However, due to computational resource limitations, we only conducted single trial for the GPT model at large dataset sizes (10,000, 20,000, and 40,000 samples). While this may introduce some variances in the results, the observed trends are consistent with expectations. A temperature of $0.5$ was used for the GPT model. }
     \label{fig: mse_data}
\end{figure*}

\textbf{Experimental Resources and Accessibility} For the training and inference of NNs, we utilize NVIDIA GTX 3090 GPUs, employing the PyTorch library to facilitate our computations. The execution of RF, KNN, and LR models is conducted on an Intel$^{\tiny{\textcircled{R}}}$ Xeon Gold$^{\tiny{\textcircled{R}}}$ 6226 CPU.
For the GPT, the entire process-from training to inference-is executed through the OpenAI API, with all computation handled by OpenAI's cloud computing.

\section{Results and Discussion}
\label{sec: experiments results}


In this section, our main focus is investigating the efficiency of FT-LLM (i.e., the fine-tuned ChatGPT 3.5 model) and comparing it with other baseline models. Additionally, we explore model performance on interpretability (Sec. \ref{learned}) and inverse design (Sec. \ref{Inverse d}).

\subsection{How efficiently does ChatGPT Learn?}

Here we investigate how efficiently LLMs can learn to predict the physical properties of materials, in terms of the quantity of training data required to achieve a certain level of prediction accuracy. On one hand, there is evidence that LLMs possess significant knowledge about natural systems, including materials, and therefore they may require less training data to make accurate predictions. On the other hand, LLMs are pre-trained on language tasks rather than numerical prediction tasks, and they also possess a massive number of free parameters; both of these properties may imply that LLMs require relatively larger training datasets to make accurate predictions.

To evaluate the learning efficiency of LLMs, we evaluate the prediction accuracy of FT-LLM against all the baseline models, as we vary the size of their training data. In Fig. \ref{fig: mse_data} we present the main results of these experiments which reveal the dependence of the two error metrics, MARE (a) and MSE (b), on the size of the training dataset. The FT-LLM achieves a best MSE of 0.00639 and a best MARE of 0.165, indicating that the FT-LLM's predictions are close to the CST-generated ground truths. This highlights the reliability of the LLM-based approach.

In Fig. \ref{fig: mse_data}(a) we find that the FT-LLM (red curve) performs similar to the linear and KNN machine learning models for the smallest dataset sizes, but shows significant improvement as the dataset size grows. At dataset sizes of 1,000 and larger, the FT-LLM outperforms all traditional machine learning models, and achieves similar accuracy to the deep neural network for all sizes thereafter.  The results therefore indicate that, within the training set sizes investigated, the LLM model achieves similar learning efficiency to the neural network. 
This result is surprising since our neural network architecture is adopted from prior work, and highly optimized for this task, and it was also trained specifically for regression, while the FT-LLM was fine-tuned using cross-entropy, which is not well-suited for regression (see Sec. \ref{sec:experimental_design_scoring_metrics}).  Despite these advantages, the FT-LLM model achieves comparable performance with the neural network after just 1,000 training samples.  

In terms of MSE, in Fig. \ref{fig: mse_data}(b) we observe that the FT-LLM performs the worst among all models in the low-data scenario ($<$ 10,000 samples). However, when larger datasets are employed, the FT-LLM shows faster improvement in its accuracy, surpassing all baseline models except for the neural network once 40,000 training samples are used.  We note that even with 40,000 training samples, the rate of improvement in the FT-LLM is greater than all other models, suggesting that it may achieve or surpass the accuracy of the neural network if more data were added.  This result is surprising since, as noted in Sec. \ref{sec:experimental_design_scoring_metrics}, all of our models were trained to minimize MSE, while the FT-LLM model was trained to minimize a different criterion.  Despite this disadvantage, the FT-LMM outperforms most models, and is improving faster than the neural network.   

The differences between the MARE and MSE evaluations can be attributed to how these two different error metrics are calculated. The spectra in our training dataset possess low numerical values (e.g. 0.001) at some frequencies. In such cases, if a prediction is 0.002 and the actual value is 0.001, the MSE is very low because it depends on the absolute error. However, MARE becomes very high in this case ($100 \%$) as it measures the relative error. LLMs that are fine-tuned using cross-entropy loss tend to maintain uniform performance across all spectral magnitudes. However, a neural network trained on MSE loss tends to achieve better accuracy in high-value regions while having lower accuracy in low-value regions because MSE does not heavily penalize small absolute errors. This explains why the FT-LLM performs competitively with neural networks in terms of MARE after 1,000 samples but requires larger dataset sizes to reach similar performance levels in terms of MSE.

We also explored the influence of the prompt template, and the LLM temperature hyperparameter, on FT-LLM's prediction accuracy. Specifically, the choice of prompt template affects how input information is structured and presented to the model, while the temperature setting controls the variability of model output.
In addition, in Appendix \ref{hardware}, we discuss the hardware efficiency of the model.

\subsubsection{Impact of Prompt} 
\begin{figure}[ptb]
    \centering
    \includegraphics[width=0.6\textwidth]{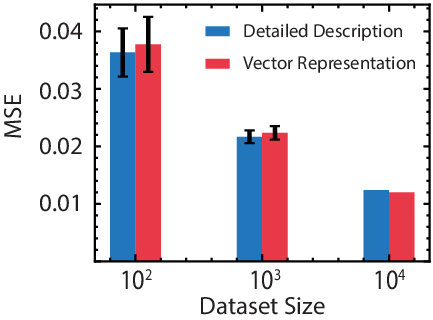}
    \caption{Evaluations of model performance using two different prompt templates. These templates are provided in Table \ref{template}. All the results presented are averages from three models. However, the results at the 10,000 data points are exceptions, as computational constraints limited these to single trials.}
    \label{fig: prompt}
\end{figure}

Studies on prompt engineering indicate that the structure and detail of the prompt may influence the performance of the LLM\cite{chen2023unleashing}. It may be conjectured that a more descriptive prompt may improve predictive accuracy. To investigate the template effect, we compared two prompts: one with relatively more detail than the other. Examples of both templates used are provided in Table \ref{template}. We conducted a fine-tuning experiment on the FT-LLM using both prompt templates with varying dataset sizes. The MSE comparison is shown in Fig. \ref{fig: prompt}. 

Our results indicate that both prompt designs yield nearly the same performance in all sizes of training samples. This consistency suggests that the accuracy of the model is not significantly influenced by whether the input data is presented in a concise vector form or as a detailed description.

\begin{table*}[!ht]
\caption{Example Prompts. These prompts are used to convert the tabular data into text sentences.}
\begin{tabular}{p{0.3\textwidth} p{0.7\textwidth}}
\toprule
\textbf{Type} & \textbf{Example} \\ \midrule
Original Data & 0.525, 1.0, 0.2, 0.188 0.1, 0.2, 0.125, 0.188 0.162, 0.162, 0.0, 0.785 -0.393, 0.0 \\ \addlinespace
Vector Representation Prompt & The All-dielectric metasurface suspend in free space is: <0.525, 1.0, 0.2, 0.188 0.1, 0.2, 0.125, 0.188 0.162, 0.162, 0.0, 0.785 -0.393, 0.0> Get the absorptivity \\ \addlinespace
Detailed Description Prompt & Calculate the absorptivity values for an all-dielectric metasurface under the following specifications: \\ &- Height: 0.525, \\ &- Periodicity: 1.0, \\ &- Elliptical resonators' dimensions and rotational angles: \\ & 1. Semi-major axis: 0.2, Semi-minor axis: 0.188, Rotation: 0.0, \\ & 2. Semi-major axis: 0.2, Semi-minor axis: 0.1, Rotation: 0.785, \\ & 3. Semi-major axis: 0.188, Semi-minor axis: 0.125, Rotation: -0.393, \\ & 4. Semi-major axis: 0.162, Semi-minor axis: 0.162, Rotation: 0.0. \\ \addlinespace
Completion & [0.004 0.005 0.006 0.007 0.008 0.01 0.012 0.015 0.018 0.022 0.028 0.038 0.054 0.101 0.384 0.309 0.358 0.496 0.354 0.198 0.135 0.13 0.152 0.189 0.238 0.297 0.372 0.465 0.534 0.566 0.585 0.597 0.62 0.652 0.587 0.532 0.544 0.595 0.642 0.684 0.632 0.552 0.541 0.55 0.568 0.62 0.65 0.637 0.633 0.664] \\
\bottomrule
\end{tabular}
\label{template}
\end{table*}

\subsubsection{Influence of Temperature}

\begin{figure}[ptb]
     \centering
     \includegraphics[width=0.6\textwidth]{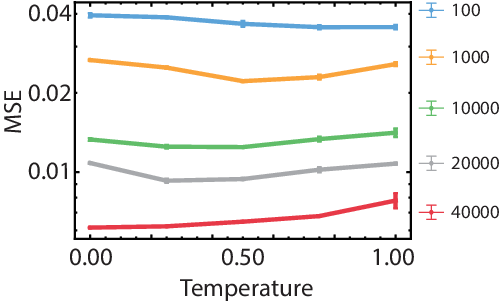}
     \caption{MSE trends for the model fine-tuned on different numbers of training samples as the temperature increases. The MSE is in $\text{log}_{10}$ scale.}
\label{fig: temp}.
\end{figure}

The temperature settings of LLMs affect the variability of their output. In general LLM's produce different responses each time it is prompted, even if the prompts are identical.  When the temperature is set lower, the LLM will tend to select a less diverse set of prompts that it estimates are most probable (i.e., the most likely to be accurate spectral predictions).  By contrast, when the temperature is higher the model will tend to provide more diverse responses, at the cost of sometimes selecting responses that it estimates are less probable. We explored the effects of varying temperature settings, specifically $[0, 0.25, 0.5, 0.75, 1]$, on model performance. The results are presented in Fig. \ref{fig: temp}. For each data size, we set different temperatures on the same models. The plot indicates that as the quantity of training data increases, the optimal temperature setting tends to be lower. We find that for small training set sizes (100 and 1000), the FT-LLM is relatively insensitive to temperature changes. Conversely, as the size of the training set increases, lower temperature settings tend to minimize the MSE. 

The results also indicate that FT-LLM produces reliable, consistent outputs. As shown in Fig. \ref{fig: temp}, in most cases, the standard deviation (std) remains at a low level across the majority of temperature settings, indicating that when presented with the same prompts multiple times, FT-LLM will consistently achieve the same level of accuracy.

\subsection{Can FT-LLM Explain What It Has Learned?}
\label{learned}
Understanding the behavior of conventional data-driven models requires additional methods, such as adding interpretability loss\cite{Wu2018} during training. By contrast, the FT-LLM is able to generate text-based responses to prompts, giving FT-LLM the potential to provide textual explanations for its predictions \cite{wei2022chain,dinh2022lift,kojima2022large}. This raises an interesting question: can the FT-LLM describe in words the physical principles underlying the metamaterial response from observation of the geometry-spectrum pairs?  To answer this question, we evaluated the FT-LLM's comprehension on electromagnetic metamaterials by asking it what the impact would be of altering various geometrical features. As an experimental control, we also provided the same responses to the un-trained ChatGPT 3.5 model, to ensure that any informative responses provided by FT-LLM were the result of training it on our data. 

Appendix \ref{interpretability} presents the questions posed and the answers generated by FT-LLM. The results indicate that the FT-LLM did not demonstrate a significantly better understanding than the original unmodified GPT model. This suggests that training on geometry-spectra pairs may not provide the model with additional understanding of the physical relationships governing our task. Furthermore, a stylistic difference was observed. The FT-LLM tends to produce responses comprising a single paragraph, in contrast to the original GPT's piecewise interpretation. This discrepancy in the output style can be attributed to the format of the training data. We have found that all completions are a continuous list of numbers, i.e. a single paragraph. It seems reasonable that this fact influenced the FT-LLM to follow this output style.

\subsection{Is FT-LLM a Good Designer?}
\label{Inverse d}
An important goal of deep learning in metamaterials is to utilize a model for spectral design through the determination of the geometry, which is called the inverse design. In comparison to regression tasks, this is a more challenging problem, which has been termed in the literature as one-to-manyness \cite{deng2021benchmarking}. Given that our FT-LLM model has demonstrated comparable performance to a deep neural network, it is of interest to evaluate its inverse performance. To do this, we gave the FT-LLM the expected spectrum and asked the model to generate a possible design. Appendix \ref{Inverse App} illustrates the example prompt and outputs.

We found that models trained on datasets that exceeded 10,000 samples were prone to producing invalid output. In the majority of instances, despite the imposition of strict constraints on the output format, our model disregarded instructions and provided merely a list of numbers. In some cases, the generated responses contained numerous repetitions and fragmented words. We tried different system prompts to address these issues, such as formatting the output structure, but the FT-LLM kept outputting meaningless completions. This may be attributed to the lack of diversity in our training dataset. As all completions are a list of numbers, this dominance in training data appears to have skewed the learning process\cite{luo2024empiricalstudycatastrophicforgetting}.  However, models fine-tuned on small data sets with strict constraints successfully generate valid geometries. We compared the inverse FT-LLM resimulation error with various inverse models including the neural adjoint\cite{deng2021neural}, tandem\cite{liu2018training,jordan2013forward}, and a conventional neural network. The results are shown in Table \ref{tab:Table1}. We find that the FT-LLM trained on a dataset size of $1,000$ achieves poor results, indicating that its designs are not accurate. 

\begin{table}[!ht]
\centering
\caption{Resimulation evaluations of inverse design algorithms with 1,000 training data. For the simulation, we use a fully connected network with $2 \times 10^{-3}$ test MSE as a surrogate model.}
\begin{tabular}{l c c}
\toprule
Algorithm &  MSE ($\times 10^{-3}$) &  MARE ($\times 10^{-1}$) \\ 
\midrule
Neural Adjoint \cite{deng2021neural} & $4.63 \pm 0.02$ & $2.24 \pm 0.018$ \\ 
Tandem \cite{liu2018training,jordan2013forward} & $17.3 \pm 0.31$ & $4.45 \pm 0.045$ \\ 
Neural Network & $15.5 \pm 0.23$  & $4.89 \pm  0.280$\\
FT-LLM & $45.7 \pm 1.77$ & $19.8 \pm  5.499$ \\ 
\bottomrule
\end{tabular}
\label{tab:Table1}
\end{table}

\section{Conclusion}

Can large-language models learn the physics of metamaterials? Our empirical study of using ChatGPT to predict the absorptivity spectrum of all-dielectric metamaterial indicates that this is indeed possible. We fine-tuned ChatGPT 3.5 with simulation data consisting of metamaterial geometry and spectra, resulting in a model termed FT-LLM. Most of the predicted spectra from the FT-LLM fine-tuned on the large dataset closely match the results from CST, an electromagnetic solver, generated ground truth, demonstrating the reliability of our approach. We also compared the FT-LLM's performance with traditional ML models, including a deep neural network that has been highly optimized for the given task. The results demonstrated that the FT-LLM achieved a comparable mean absolute relative error (MARE) to the deep neural network, once the training dataset exceeded a size of 1,000. In terms of MSE, the FT-LLM had poor performance for small training dataset sizes. However, its performance increased substantially and faster than other ML models with increasing training dataset size. This capacity to process and analyze large datasets demonstrates that LLMs can serve as surrogate models in the electromagnetics research, exhibiting significant potential for predictive metamaterial physics. 

On the other hand, interpretability and inverse design remain challenges. Although the FT-LLM model could generate reasonable answers related to physics, it did not show any better understanding of our particular problem, even after fine-tuning on a large dataset of related data. Additionally, using the FT-LLM for inverse design proved ineffective, as traditional inverse design methods achieved better results. The shortcomings exhibited by the FT-LLM studied here indicates the need for further research.

The results of this study have broader implications for the application of AI in the physical sciences and engineering disciplines. By integrating different types of data, foundational models can provide a more holistic understanding of scientific phenomena, uncovering hidden patterns, and providing physical insights. This adaptability makes LLMs valuable tools in fields such as optics, photonics, and materials science, where they can accelerate progress and faster interdisciplinary research.

\textbf{Future work:}  Our work tested two different geometry representation methods, but both of them are specific to the dataset.  Inspired by the success of SMILES \cite{weininger1988smiles} in chemistry (a technique that uses linear text to describe the structure of chemical species), we propose that the development of a standardized encoding scheme designed for representing the geometry of all metamaterials could accelerate LLMs research in this field.

In addition, we encountered problems with poor interpretability and inverse design capabilities. This may be because our dataset only includes the forward prediction question-answer sets. In this case, creating a more diverse dataset, including related research papers and textbooks, may help address these issues.

Another promising direction is building LLM agent systems. LLMs have proven effective in the extraction of knowledge from papers \cite{circi2023retrieval}, and automating experimental procedures \cite{hu2024language,boiko2023autonomous}, while our work highlights their potential in predicting metamaterial physics. Building an agent system that could autonomously interact with complex datasets, search for information in the paper, and analyze the results is feasible. In metamaterial research, such LLM agents could enhance productivity and innovation.

\bibliographystyle{IEEEtran}  
\bibliography{references}  

\newpage

\appendix 
\section*{\centering Appendix}
\section{Managing Output Length and Conversion Errors}
\label{A: Managing output errors}
One challenge with evaluating our LLMs is that, by design, they do not produce fixed-length sequences, and they can add extra unnecessary characters to their predictions (e.g., commas, white space, punctuation marks).  The length of the ground truth spectra $\text{Len} (S_t)$ is $50$.  However, $\text{Len} (S_p)$ generated by FT-LLM, may differ from $50$.  We have found that for training sizes of $10,000$ and $1,000$ FT-LLM provides outputs of length $\text{Len} (S_p)=50$, 98\% and 85\% of the time, respectively. For smaller and larger training data sets, the output began to have more occurrences of $\text{Len} (S_p)=49$ or $\text{Len} (S_p)=51$, respectively. Our process for dealing with these instances is that we regenerate the output up to 3 times to attempt to achieve $L_t=L_p$, otherwise we will only compare $S_t[0:L_{min}]$ and $S_p[0:L_{min}]$, where $L_{min} = \text{min}(\text{Len} (S_t), \text{Len} (S_p))$.

Another issue that may arise is the conversion of the text output from the LLM into numerical values. For example, a text output of "0.0.13" cannot be converted to a numeric value. In the case that, the output is regenerated and the process is repeated up to 3 times. If all attempts fail, we then skip this data point while evaluating the base-line models and FT-LLM. We have found that our FT-LLM outputs non-numerical data less than 0.1\% of the time. Additionally, even if the numerical transformation is successful, errors can still occur. For example, the completion output "0 103" could be transferred to numbers 0 and 13, but it should represent "0.103," which misses a decimal point. Given that our data falls within the range of 0 to 1, any completions that do not comply with this format are regarded as errors.

\section{ Hyperpaprameter Tuning}

\label{A:Hyperpara}
To optimize hyperparameters, we used Bayesian optimization via the Scikit-Optimize package \cite{https://doi.org/10.5281/zenodo.1207017}. We employed the default settings in Scikit-Optimize, conducting 30 Bayesian trials for each traditional model. We used this approach to efficiently explore the parameter space to identify optimized configurations based on the validation MSE\cite{Victoria2020}. For the neural network, we used the same setting in the previous research \cite{deng2021neural} and only changed the output dimensions. We focus on fine-tuning the following models:

\begin{itemize}
    \item Random Forest (RF):
    \begin{itemize}
        \item Number of trees: 33
        \item Minimum samples to split an internal node: 4
        \item Minimum samples to be at a leaf node: 2
        \item Maximum depth: 14
    \end{itemize}
    
    \item Neural Network (NN):
    \begin{itemize}
        \item Learning Rate: 1e-3
        \item Number of Layers: 10
        \item Number of Neurons in each layer: 2000
        \item Activation Function: ReLu
        \item Optimizer: Adam
    \end{itemize}
    
    \item K-Nearest Neighbors (KNN):
    \begin{itemize}
        \item Number of neighbors: 12
    \end{itemize}

        \item Linear Regression (LR):
    \begin{itemize}
        \item We used the default setting of the Sklearn\cite{sklearn_api}.
    \end{itemize}
\end{itemize}

\section{ Handling Division by Zero in MARE Calculation}
\label{MARE}
In the test dataset, one spectrum has zero value data. In the MARE calculation, we need to divide the ground truth, but dividing by zero is invalid. To deal with this, a common strategy is to add a small constant to the denominator\cite{Kim2016}. Given that the smallest non-zero value in the dataset is 0.001, we chose to add 0.0001 in our study. This effectively handles the division without significantly affecting the overall results.

\section{ Model Hardware Efficiency}
\label{hardware}
computation usage is an important aspect while evaluating model performance\cite{Choupanzadeh2023-tz}. How efficiently does FT-LLM utilize computational resources? In this study, we evaluate training time and memory usage in various baseline models and the FT-LLM. Table \ref{tab:memory_time} shows the results. 

Fine-tuning ChatGPT requires significantly more computational time due to its larger model size. As the fine-tuning process was run on the OpenAI platform, we were limited by platform constraints that were beyond our direct control. Future work focusing on reducing this computational time could include using quantized models which lower the numerical precision of model parameters while preserving high prediction accuracy\cite{MLSYS2024_5edb57c0} and exploring more efficient fine-tuning techniques.

In terms of memory, the fine-tuning of ChatGPT was conducted on OpenAI's platform so that we did not need to do the local computation, which is good when the resources are limited. However, the total price for doing all the FT-LLM experiments in the paper on OpenAI's platform was approximately \$4,000. This indicates the extremely high computational resources required by LLMs.

\begin{table}[!ht]
\centering
\caption{Training time and memory usage for each model on a dataset of 1,000 samples. Memory usage for FT-LLM is not available.}
\begin{tabular}{l c c}
\toprule
Algorithm &  Time (s) &  Computational Resources \\ 
\midrule
Linear Regression & $0.79 $ & $16.3 $ MB \\ 
Random Forest & $0.73 $ & $13.9 $ MB\\ 
K Nearest Neighbors & $0.89 $ & $13.8$ MB\\ 
Feed-forward Neural Networks & $67.2 $  & $1,083$ MB\\
FT-LLM & $2753$  & NA\\ \bottomrule
\end{tabular}
\label{tab:memory_time}
\end{table}

\section{ Examples of Inverse Design}
\label{Inverse App}

\subsection{Example Outputs}
\label{Inverse designs}
The following demonstrates the outputs of a ChatGPT model fine-tuned with different numbers of training samples. The prompts represent an inverse design task for specifying a metamaterial with a
desired spectrum. The completion of the model fine-tuned with 1,000 samples yields a physically implausible design, since the axes of
resonators are greater than half of the periodicity. In addition, the outputs of the model fine-tuned with 10,000 samples are invalid.
Temperature is set as 0.5.

\textbf{Input Prompt:}

You are an electromagnetic expert. You can design a metamaterial with a given desired spectrum. Your output should fill the [value] in the following:

An all-dielectric metasurface under the following specifications:

- Height: [value]
    
- Periodicity: [value]
    
- Elliptical resonators' dimensions and rotational angles:
    
- 1. Semi-major axis: [value], Semi-minor axis: [value], Rotation: [value]
    
 2. Semi-major axis: [value], Semi-minor axis: [value], Rotation: [value]
    
- 3. Semi-major axis: [value], Semi-minor axis: [value], Rotation: [value]
    
- 4. Semi-major axis: [value], Semi-minor axis: [value], Rotation: [value]
    
The desired absorptivity values should have a peak value (greater than 0.8) in the 40th values.

\textbf{Completion from model fine-tuned in 1000 training
samples:}

an all-dielectric metasurface under the following specifications:

Height: 500,

Periodicity: 800,

Elliptical resonators’ dimensions and rotational angles:

1. Semi-major axis: 300, Semi-minor axis: 150, Rotation: 0,

2. Semi-major axis: 200, Semi-minor axis: 100, Rotation: 0,

3. Semi-major axis: 150, Semi-minor axis: 75, Rotation: 0,

4. Semi-major axis: 150, Semi-minor axis: 75, Rotation: 0

\textbf{Completion 1 from model fine-tuned in 10,000 training samples:}

[0.006 0.008 0.011 0.014 0.017 0.021 0.026 0.032 0.04 0.059 0.12 0.141 0.092 0.1 0.126 0.154 0.197 0.252 0.264 0.233 0.247 0.294 0.354 0.409 0.466 0.499 0.502 0.484 0.484 0.513 0.556 0.647 0.621 0.553 0.519 0.515 0.535 0.563 0.585 0.595 0.587 0.574 0.563 0.551 0.536 0.521 0.509 0.5 0.497 0.499]

\textbf{Completion 2 from model fine-tuned in 10,000 training samples:}

Infeasible. No such solution.@@No such design@@. @@No such design@@. @@No such design@@. @@No such design@@. @@No such design@@. @@No such design@@. @@No such design@@. @@No such design@@. @@No such design@@. @@No such design@@. @@No such design@@. @@No such design@@. @@No such design@@. @@No such design@@. @@No such design@@. @@23@@18@@16@@18@@21@@26@@32@@41@@ 52@@63@@65@@67@@69@@74@@79@@84@@86@ @88@@81@@8.2@@78@@73@@74@@75@@81@@7 8@@68@@62@@62@@69@@79@@72@@66@@66@

\subsection{Example Prompt}
As demonstrated in the Appendix \ref{Inverse designs}, when the model is trained on a dataset exceeding 10,000 instances, it produces invalid outputs. We tried to use different prompts to address this issue, but the model remained unable to generate the desired output. The prompts used are listed below.

\begin{itemize}
    \item You are a metamaterial expert. Please design an all-dielectric metasurface suspended in free space whose absorptivity values have a peak value (greater than 0.8) of 400 THZ. It would help if you could give me the common height, periodicity, semi-major and semi-major axis of four elliptical resonators, and the angles of rotation with respect to a vertical axis.
    \item You are a ChatGPT model fine-tuned on metamaterial simulation data, enabling you to perform inverse design. Your task is to design an all-dielectric metasurface suspended in free space that achieves a peak absorptivity greater than 0.8 at 400 THz. Please provide the following parameters in the order given: [height, periodicity, semi-major axis and semi-major axis of the four elliptical resonators, Rotational angles for each resonator].
    \item You are a ChatGPT model trained for inverse design. Design an all-dielectric metasurface suspended in free space, achieving a peak absorptivity greater than 0.8 at 400 THz.
    Important: Your output must only contain the values: Height,periodicity, semi-major and semi-minor axes of four elliptical resonators, rotational angles of four elliptical resonators. Random or irrelevant outputs such as lists of decimals or strings are not acceptable.

\end{itemize}

\section{ Examples of model completions for explaining feature importance}
\label{interpretability}

This section demonstrates examples of model completions for explaining feature importance. original LLM refers to the GPT 3.5-1106 without fine-tuning. On the other hand, the fine-tuned version of GPT 3.5 has been trained on a dataset of 10,000 data points, using a detailed description template. The temperature is set to 1.

\subsection{Heights Influence}
\textbf{Input Prompt:}

You are a physics expert. What is the influence
of changing heights on the absorptivity of an all-
dielectric metasurface?

\textbf{Completion of FT-LLM:}

Changing the height of the dielectric metasurface affects the resonant modes and the overall absorption. As the height increases,the resonant modes shift to longer wavelengths, and the absorption at those wavelengths increases. This is because the increased height allows for more interaction between the incident electromagnetic wave and the metasurface, leading to stronger scattering and absorption. At the same time, the absorption at non-resonant wavelengths may decrease as the height increases, due to reduced interaction with the metasurface. However, the overall influence of changing heights on the absorptivity depends on the specific design and parameters of the metasurface.

\textbf{Completion of Original LLM:}

Changing the height of an all-dielectric metasurface can have a significant influence on its absorptivity. The absorptivity of a metasurface refers to its ability to absorb incident electromagnetic radiation. 

By changing the height of the metasurface, the resonant modes of the structure can be tuned, which in turn affects its absorptivity. For example, changing the height can shift the resonant frequency of the metasurface, leading to changes in the absorption properties at specific frequencies. Additionally, changing the height can also affect the coupling between the metasurface and the incident electromagnetic radiation, which can further influence the overall absorptivity of the structure. 
 
In general, tuning the height of an all-dielectric metasurface can provide a degree of control over its absorptivity. This allows for custom design and optimization for specific applications such as solar cells, sensors, and stealth technology.

\subsection{periodicity Influence}
\textbf{Input Prompt:}

You are a physics expert. What is the influence
of changing periodicity on the absorptivity of an all-
dielectric metasurface?

\textbf{Completion of FT-LLM:}

Varying the periodicity of an all-dielectric metasurface can significantly influence its absorptivity. As you change the periodicity, the effective refractive index can be altered, leading to different Mie resonances and interference effects. This can result in modified absorption spectra and different dispersion characteristics. In particular, as the periodicity is modified, the coupling between the metasurface and the incident radiation can change, affecting the intensity of the fields around and within the metasurface, consequently influencing the absorption. Additionally, the characteristic Mie resonances in the absorptivity can change with the periodicity due to the modified phase matching conditions, leading to different absorption features. Therefore, it's important to carefully consider the impact of periodicity on the design of all-dielectric metasurfaces in order to achieve the desired optical properties.

\textbf{Completion of Original LLM:}

The absorptivity of an all-dielectric metasurface can be influenced by changing the periodicity of the structure in several ways: 

1. Bandwidth of absorption: Changing the periodicity can affect the resonance of the metasurface and thereby change the range of wavelengths over which absorption occurs. By adjusting the periodicity, one can tune the absorptivity peak to different wavelengths, allowing for control over the bandwidth of absorption. 

2. Absorption efficiency: The efficiency of absorption can also be affected by changing the periodicity. Optimizing the periodicity can enhance the absorption efficiency by matching the resonance condition to the incident light wavelength more closely. 

3. Polarization sensitivity: The absorptivity of a metasurface can also be influenced by the polarization of the incident light. Changing the periodicity can lead to different responses for different polarizations, allowing for polarization-selective absorption. 

4. Angular dependence: The absorptivity of a metasurface can exhibit angular dependency, meaning that the absorption properties change with the angle of incidence of the light. By tuning the periodicity, one can control the angular dependence of absorption. 

Overall, changing the periodicity of an all-dielectric metasurface provides a versatile way to tailor its absorptivity properties for specific applications, allowing for optimization of absorption efficiency, bandwidth, and other characteristics.

\subsection{elliptical resonators Influence}

\textbf{Input Prompt:}

You are a physics expert. What is the influence of changing the angle of the elliptical resonators on the absorptivity of an all-dielectric metasurface?

\textbf{Completion of FT-LLM}

By changing the angle of elliptical resonators in an all-dielectric metasurface, you change their orientation w.r.t incident electric field. This, in turn, will change the effective refractive index and the support of different Mie-type electric resonances. This will influence the energy distribution and the total amount of absorbed energy at the different resonant modes, and thus effectively change the absorptivity of the metasurface. You can expect changes in the resonant amplitudes and redshift or blueshift of resonance due to the rotation, all contributing to the alteration of the absorptive properties. It is similar to rotate a polarizer, the absorptive maxima dip and flip depending on the geometry and orientation of the polarization w.r.t the axes of the film. For example, the periodic array of elliptical resonators will have periodicity, orientation, and asymmetry, thus the absorption is dictated by the resonant response which in turn is affected by the orientation effects. The sensitive dependency of the rotation on the response means careful consideration of the material orthorhombic nature will have to be made for precise and accurate characterization. The resulting informatics will tell engineers and researchers how much the geometry must be rotated to extract the best performance or to keep the performance consistent to create an array of the particular geometry with lessening reliance on the variation at the different polarization angles. For different resonant orientations the metasurfaces whose symmetries will dictate absorption pattern into forward and backward scattering, thus understanding the orientations can aid in possible secondary uses or in reducing backscattering losses or optimizing forward scattering collimating emission.

\textbf{Completion of Original LLM:}

Changing the angle of elliptical resonators in an all-dielectric metasurface can have a significant impact on the absorptivity of the metasurface. The absorptivity of a metasurface is determined by its ability to efficiently absorb incident electromagnetic energy. This ability is closely related to the resonant behavior of the individual resonators within the metasurface. 

Elliptical resonators exhibit different electromagnetic responses compared to circular or rectangular resonators, due to their asymmetric geometry. By changing the angle of the elliptical resonators, you can tune the resonant frequency of the individual resonators within the metasurface. This tuning can lead to enhanced absorption properties at specific frequencies or polarization states. 

In general, adjusting the angle of elliptical resonators can affect the overall response of the metasurface by influencing the coupling between adjacent resonators, altering the effective refractive index of the metasurface, and modifying the scattering properties of the structure. These changes can result in a tailored absorptivity spectrum for specific applications such as sensing, energy harvesting, or stealth technology. 

Therefore, careful design and optimization of the geometry and orientation of elliptical resonators in an all-dielectric metasurface can provide control over the absorptivity of the metasurface and enable customization of its electromagnetic response for desired functionalities.

\subsection{radius Influence}

\textbf{Input Prompt:}

You are a physics expert. What is the influence of changing the radius of the elliptical resonators on the absorptivity of an all-dielectric metasurface?

\textbf{Completion of FT-LLM:}

As the radius of the resonator is changed, the absorptivity of the metasurface varies accordingly. For smaller radii, the absorptivity can increase significantly in the shorter wavelengths due to the presence of electric dipolar resonances. However, at larger radii, the electric dipolar resonance shifts to longer wavelengths and diminishes, while magnetic dipolar resonances arise leading to high absorption in the longer wavelengths. Overall, the absorptivity of the metasurface can be tuned by changing the radius, making it possible to design devices with specific absorption spectra.

\textbf{Completion of Original LLM:}

Changing the radius of elliptical resonators in an all-dielectric metasurface can have a significant impact on the absorptivity of the metasurface. The absorptivity of a metasurface is determined by its ability to effectively absorb incident electromagnetic radiation. 

In the case of elliptical resonators, the resonant behavior is highly dependent on the size and shape of the resonators. By changing the radius of the elliptical resonators, the resonance frequency of the resonators can be tuned. This allows for control over the absorption properties of the metasurface. 

Increasing the radius of the elliptical resonators can lead to a redshift in the resonance frequency, resulting in enhanced absorption at longer wavelengths. Conversely, decreasing the radius can lead to a blueshift in the resonance frequency, enhancing absorption at shorter wavelengths. 

Therefore, adjusting the radius of the elliptical resonators in an all-dielectric metasurface can be a powerful way to tailor its absorptivity and optimize its performance for specific applications such as solar energy harvesting, thermal emission control, or sensing.

\end{document}